\documentclass[12pt]{article}

\begin{document}

\begin{center}
{\Large\bf{}Modification of the Bel-Robinson type energy-momentum}
\end{center}

\begin{center}
Lau Loi So\footnote{email address: s0242010@gmail.com}\\
Department of Physics, National Central University, Chung-Li 320,
Taiwan
\end{center}

\begin{abstract}
For describing the non-negative gravitational energy-momentum in
terms of a pure Bel-Robinson type energy-momentum in a quasilocal
2-surface, both the Bel-Robinson tensor $B$ and tensor $V$ are
suitable. We found that this Bel-Robinson type energy-momentum can
be modified such that it satisfies the Lorentz covariant, future
pointing and non-spacelike properties. We find that these
particular energy-momentum properties can be obtained from (i):
$B$ or $V$ plus a tensor $S$ in a quasilocal small cube limit, or
(ii): directly evaluating the energy-momentum of $B$ or $V$ in a
quasilocal small box region.
\end{abstract}

\section{Introduction}
According to the Living Review article, Szabados (see 4.2.2
in~\cite{Szabados}) argued that the natural choice and the only
choice for describing the non-negative gravitational quasilocal
energy-momentum expression is the tensor that possesses the
Bel-Robinson type energy-momentum. This energy-momentum density
satisfies the Lorentz covariant, future pointing and non-spacelike
properties.  The Bel-Robinson tensor $B$ and tensor
$V$~\cite{SoCQG2009} both fulfil these requirements. They are
defined in empty space as follows:
\begin{eqnarray}
B_{\alpha\beta\xi\kappa}&:=&R_{\alpha\lambda\xi\sigma}R_{\beta}{}^{\lambda}{}_{\kappa}{}^{\sigma}
+R_{\alpha\lambda\kappa\sigma}R_{\beta}{}^{\lambda}{}_{\xi}{}^{\sigma}
-\frac{1}{8}g_{\alpha\beta}g_{\xi\kappa}\mathbf{R}^{2},\\
V_{\alpha\beta\xi\kappa}&:=&R_{\alpha\xi\lambda\sigma}R_{\beta\kappa}{}^{\lambda\sigma}
+R_{\alpha\kappa\lambda\sigma}R_{\beta\xi}{}^{\lambda\sigma}
+R_{\alpha\lambda\beta\sigma}R_{\xi}{}^{\lambda}{}_{\kappa}{}^{\sigma}
+R_{\alpha\lambda\beta\sigma}R_{\kappa}{}^{\lambda}{}_{\xi}{}^{\sigma}
-\frac{1}{8}g_{\alpha\beta}g_{\xi\kappa}\mathbf{R}^{2},\quad
\end{eqnarray}
where $R_{\alpha\beta\xi\kappa}$ is the Riemann curvature,
$\mathbf{R}^{2}=R_{\rho\tau\xi\kappa}R^{\rho\tau\xi\kappa}$, Greek
letters mean spacetime and the signature we use is $+2$. The
associated energy-momentum values are
\begin{equation}
B_{\mu{}0ij}\delta^{ij}\equiv{}V_{\mu{}0ij}\delta^{ij}
=(E_{ab}E^{ab}+H_{ab}H^{ab},2\epsilon_{cab}E^{a}{}_{d}H^{bd}),\label{29aSep2013}
\end{equation}
where Latin denotes spatial indices. The electric part $E_{ab}$
and magnetic part $H_{ab}$, are defined in terms of the Weyl
curvature~\cite{Carmeli}: $E_{ab}:=C_{ambn}t^{m}t^{n}$ and
$H_{ab}:=*C_{ambn}t^{m}t^{n}$, where $t^{m}$ is the timelike unit
vector and $*C_{\rho\tau\xi\kappa}$ indicates its dual for the
evaluation.  Here we emphasize that the energy component in
(\ref{29aSep2013}) is non-negative which is well known and the
momentum component is a kind of cross product between $E$ and $H$:
\begin{eqnarray}
\epsilon_{cab}E^{a}{}_{d}H^{bd}&=&E_{ab}\times{}H^{ab}\nonumber\\
&=&E_{1a}\times{}H^{1a}+E_{2a}\times{}H^{2a}+E_{3a}\times{}H^{3a}\nonumber\\
&=&|E_{1a}||H_{1b}|\sin\theta_{1}+|E_{2a}||H_{2b}|\sin\theta_{2}+|E_{3a}||H_{3b}|\sin\theta_{3},
\end{eqnarray}
where $\theta_{1}$ is the angle between $E_{1a}$ and $H_{1a}$,
similarly for $\theta_{2}$ and $\theta_{3}$.

It is known that if the quasilocal energy-momentum expression is
positive on a large scale (e.g., ADM mass at spatial
infinity,~\cite{SoCQG2008}) this ensures positive energy in the
small region limit (e.g., $B_{00ij}\delta^{ij}$~\cite{Szabados}).
Even though the converse is not true, however, negative energy in
a small region guarantees negative on the large. The positivity
energy proof is not easy, but one can use the quasilocal
energy-momentum expression in a small 2-surface as a simple test.
We examined that only $B$ and $V$ produce the Bel-Robinson type
energy-momentum in a quasilocal small
region~\cite{CQG262009085004}. Previously, we believed that only
the Bel-Robinson type energy-momentum satisfy the Lorentz
covariant and future directed non-spacelike properties. Now we
claim this is not the case. Moreover, after discovering how to
modify the Bel-Robinson type energy-momentum, we know how to
choose a proper quasilocal energy-momentum expression that
produces the desired energy-momentum.

Referring to (\ref{29aSep2013}), both $B$ and $V$ have the same
Bel-Robinson type energy-momentum which has the desired
energy-momentum relationship, i.e., causal:
\begin{equation}
(E_{ab}E^{ab}+H_{ab}H^{ab})-|2\epsilon_{cab}E^{a}{}_{d}H^{bd}|\geq{}0.\label{9aOct2013}
\end{equation}
Here we consider two more possibilities for the comparison with
the energy and still obtain the non-negative condition:
\begin{eqnarray}
(E_{ab}E^{ab}+H_{ab}H^{ab})+k_{1}(E_{ab}E^{ab}-H_{ab}H^{ab})\geq0,\quad{}&\Rightarrow&\quad{}
|k_{1}|\leq1,\label{22aOct2013}\\
(E_{ab}E^{ab}+H_{ab}H^{ab})+k_{2}E_{ab}H^{ab}\geq0,\quad{}&\Rightarrow&\quad{}
|k_{2}|\leq2.\label{22bOct2013}
\end{eqnarray}
The above two extra invariant terms come from
\begin{eqnarray}
R_{\alpha\beta\mu\nu}R^{\alpha\beta\mu\nu}=8(E_{ab}E^{ab}-H_{ab}H^{ab}),\quad{}
R_{\alpha\beta\mu\nu}*R^{\alpha\beta\mu\nu}=16E_{ab}H^{ab}.
\end{eqnarray}
The first term can be classified as the energy density and the
second as the momentum density.  The momentum density can be
classified as a dot product between $E$ and $H$
\begin{eqnarray}
E_{ab}H^{ab}&=&E_{1a}H^{1a}+E_{2a}H^{2a}+E_{3a}H^{3a}\nonumber\\
&=&|E_{1a}||H_{1b}|\cos\theta_{1}+|E_{2a}||H_{2b}|\cos\theta_{2}+|E_{3a}||H_{3b}|\cos\theta_{3},
\end{eqnarray}
Combining the inequalities from (\ref{9aOct2013}) to
(\ref{22bOct2013})
\begin{equation}
(E_{ab}E^{ab}+H_{ab}H^{ab})+k_{1}(E_{ab}E^{ab}-H_{ab}H^{ab})
+k_{2}E_{ab}H^{ab}-|2\epsilon_{cab}E^{a}{}_{d}H^{bd}|
\geq{}0.\label{27aSep2013}
\end{equation}
According to~\cite{Szabados}, the above non-negative inequality
holds only if $k_{1}$ and $k_{2}$ are zero.  However, we disagree
with this statement.  Before we proceed, we will here demonstrate
analogous behaviour using electromagnetism. Since gravity allows
this kind of modification, electromagnetism should have a similar
interesting result. Let the electromagnetic energy-momentum tensor
in Minkowski coordinates be
\begin{eqnarray}
T_{\alpha\beta}:=F_{\alpha\lambda}F_{\beta}{}^{\lambda}-\frac{1}{4}g_{\alpha\beta}F_{\lambda\sigma}F^{\lambda\sigma},
\end{eqnarray}
where $F_{\mu\nu}$ is the electromagnetic field tensor. The known
relationship for the energy and momentum is
\begin{equation}
T_{00}-|T_{0c}|=\frac{1}{2}(|\vec{E}|^{2}+|\vec{B}|^{2})-|\vec{E}\times\vec{B}|\geq0.\label{8aOct2013}
\end{equation}
Here we modify (\ref{8aOct2013}) by adding two more invariants
from the quadratic of electromagnetic field tensor
\begin{eqnarray}
F_{\alpha\beta}F^{\alpha\beta}=-2(|\vec{E}|^{2}-|\vec{B}|^{2}),\quad{}
F_{\alpha\beta}*F^{\alpha\beta}=4(\vec{E}\cdot\vec{B}).
\end{eqnarray}
Assuming the magnitude of the electric field $|\vec{E}|$ and
magnetic field $|\vec{B}|$ is linearly proportional:
$|\vec{B}|=\alpha|\vec{E}|$ and $\alpha\geq0$. Particularly,
consider the following combination
\begin{eqnarray}
&&(|\vec{E}|^{2}+|\vec{B}|^{2})+\delta_{1}(|\vec{E}|^{2}-|\vec{B}|^{2})
+\delta_{2}\vec{E}\cdot\vec{B}-2|\vec{E}\times\vec{B}|\nonumber\\
&=&(|\vec{E}|^{2}+|\vec{B}|^{2})+\delta_{1}(|\vec{E}|^{2}-|\vec{B}|^{2})
+\delta_{2}|\vec{E}||\vec{B}|\cos\varphi-2|\vec{E}||\vec{B}||\sin\varphi|\nonumber\\
&\geq&(1+\alpha^{2})|\vec{E}|^{2}+\delta_{1}(1-\alpha^{2})|\vec{E}|^{2}
-|\delta_{2}|\alpha|\vec{E}|^{2}|\cos\varphi|
-2\alpha|\vec{E}|^{2}|\sin\varphi|\nonumber\\
&=&\left\{(1-\alpha)^{2}\left[1+\frac{\delta_{1}(1+\alpha)}{(1-\alpha)}\right]
+2\alpha\left(1-\frac{1}{2}|\delta_{2}||\cos\varphi|-|\sin\varphi|\right)\right\}|\vec{E}|^{2}\nonumber\\
&\geq&0,
\end{eqnarray}
provided that
\begin{equation}
\delta_{1}\geq\frac{(\alpha-1)}{(\alpha+1)},\quad{}
|\delta_{2}|\leq\frac{2(1-|\sin\varphi|)}{|\cos\varphi|},
\end{equation}
where $\varphi$ is the angle between $\vec{E}$ and $\vec{B}$.
Apply the above method to general relativity, let
$|H_{Ia}|=\alpha_{I}|E_{Ia}|$ and $\alpha_{I}\geq0$, where
$I=1,2,3$ and it is not sum, consider (\ref{27aSep2013})
\begin{eqnarray}
&&(E_{ab}E^{ab}+H_{ab}H^{ab})+k_{1}(E_{ab}E^{ab}-H_{ab}H^{ab})+k_{2}E_{ab}H^{ab}-2|\epsilon_{cab}E^{ad}H^{b}{}_{d}|\nonumber\\
&\geq&(E^{2}_{ab}+H^{2}_{ab})+k_{1}(E^{2}_{ab}-H^{2}_{ab})-|k_{2}||E_{ab}H^{ab}|
-2|E_{ab}\times{}H^{ab}|\nonumber\\
&\geq&(1+k_{1})E^{2}_{1a}+(1-k_{1})H^{2}_{1a}
-|k_{2}||E_{1a}||H_{1b}||\cos\theta_{1}|-2|E_{1a}||H_{1b}||\sin\theta_{1}|
\nonumber\\
&&+(1+k_{1})E^{2}_{2a}+(1-k_{1})H^{2}_{2a}
-|k_{2}||E_{2a}||H_{2b}||\cos\theta_{2}|-2|E_{2a}||H_{2b}||\sin\theta_{2}|\nonumber\\
&&+(1+k_{1})E^{2}_{3a}+(1-k_{1})H^{2}_{3a}
-|k_{2}||E_{3a}||H_{3b}||\cos\theta_{3}|-2|E_{3a}||H_{3b}||\sin\theta_{3}|\nonumber\\
&=&\left\{ (1-\alpha_{I})^{2}\left[1
+\frac{k_{1}(1+\alpha_{I})}{(1-\alpha_{I})}\right]+2\alpha_{I}\left(1-\frac{1}{2}|k_{2}||\cos\theta_{I}|
-|\sin\theta_{I}|\right) \right\}E^{2}_{Ia}
\nonumber\\
&\geq&0, \label{3aDec2013}
\end{eqnarray}
assumed that
\begin{eqnarray}
k_{1}\geq\frac{(\alpha_{I}-1)}{(\alpha_{I}+1)},\quad{}|k_{2}|\leq\frac{2(1-|\sin\theta_{I}|)}{|\cos\theta_{I}|}.
\end{eqnarray}
Indeed (\ref{3aDec2013}) is non-negative for some non-vanishing
$k_{1}$ and $k_{2}$.  However, our result is strictly forbidden
according to the conclusion of Szabados's article~\cite{Szabados}.

Here comes a natural question: If (\ref{3aDec2013}) is correct,
what are the exact ranges for $k_{1}$ and $k_{2}$? More precisely,
looking at (\ref{27aSep2013}) again, we reexamine what are the
ranges for constants $k_{1}$ and $k_{2}$ that can be selected such
that the Lorentz covariant and future directed non-spacelike
qualities are not altered.  For this purpose we use the 5 Petrov
types~\cite{Lobo} Riemann curvature for the verification. After
some simple algebra, we find a different results from
Szabados~\cite{Szabados}:
\begin{equation} |k_{1}|\leq1, \quad{}
|k_{2}|\leq2(1-|k_{1}|).\label{3bDec2013}
\end{equation}
This indicates that, in terms of a quasilocal energy-momentum
expression, $B$ and $V$ are not the only candidate that satisfy
the Lorentz covariant and future directed non-spacelike
requirements. There exists some relaxation freedom for the
modification. (i): $B$ or $V$ plus a tensor
$S_{\alpha\beta\xi\kappa}=R_{\alpha\xi\lambda\sigma}R_{\beta\kappa}{}^{\lambda\sigma}
+R_{\alpha\kappa\lambda\sigma}R_{\beta\xi}{}^{\lambda\sigma}
+\frac{1}{4}g_{\alpha\beta}g_{\xi\kappa}\mathbf{R}^{2}$ in a
quasilocal small cube limit or (ii): directly evaluate the
energy-momentum for $B$ or $V$ in a quasilocal small box region.

\section{Quasilocal energy-momentum}
We examine the gravitational quasilocal energy-momentum, which
satisfies the Lorentz covariant and future directed non-spacelike
conditions by two approaches.

Case (i): Consider a simple physical situation such that within a
small cube limit we define: $\mathbf{t}+sS$, where $\mathbf{t}$
can be replaced by $B$ or $V$, and $s$ is a constant. For constant
time $t_{0}=0$, the energy-momentum in vacuum with a finite
dimension $a_{0}$
\begin{equation}
P_{\mu}=\int_{t_{0}}(\mathbf{t}^{0}{}_{\mu\xi\kappa}+sS^{0}{}_{\mu\xi\kappa})x^{\xi}x^{\kappa}dV
=\frac{1}{12}a^{5}_{0}(\mathbf{t}^{0}{}_{\mu{}ij}+sS^{0}{}_{\mu{}ij})\delta^{ij}.\label{5aOct2013}
\end{equation}
Based on~\cite{Szabados}, the only possibility is $s=0$ in order
to produce the Lorentz covariant, future pointing and
non-spacelike properties. However, we claim that there are some
$s\neq0$ such that these properties are preserved. As the
4-momentum of
$S_{0\mu{}ij}\delta^{ij}=-10(E^{2}_{ab}-H^{2}_{ab},0,0,0)$, we
only vary the energy without affecting the momentum. After the
substitution, the energy for (\ref{5aOct2013}) is
\begin{equation}
E=\frac{1}{12}a^{5}_{0}\left[(E_{ab}E^{ab}+H_{ab}H^{ab})-10s(E_{ab}E^{ab}-H_{ab}H^{ab})\right],\label{19aOct2013}
\end{equation}
and the associated momentum is
$P_{c}=\frac{1}{12}a^{5}_{0}(2\epsilon_{cab}E^{a}{}_{d}H^{bd})$.
The criterion for non-negative energy in (\ref{19aOct2013}) is
that $|s|\leq{}1/10$. However, since the values of $E_{ab}$ and
$H_{ab}$ can be arbitrary at a given point, obviously the sign of
the energy component of $S$ is uncertain. The outcome is that $S$
affects the desired Bel-Robinson type energy-momentum inequality:
$E\geq|\vec{P}|$. Previously, our preference was achieving a
multiple of pure Bel-Robinson type energy-momentum in a small
sphere or box~\cite{SoCQG2009,Garecki}, and we thought the result
in (\ref{5aOct2013}) was strictly forbidden unless $s=0$. However,
we now claim that this is not true: we find a certain linear
combinations between $\mathbf{t}$ and $S$ that are legitimate.
Comparing (\ref{3aDec2013}) and (\ref{19aOct2013}), we observed
that $k_{1}=s\leq1/10\leq1$ and $k_{2}=0$. Indeed it does satisfy
the Lorentz covariant and future directed non-spacelike
requirements.

Case (ii): Demonstrate the energy-momentum in a small box for
replacing $\mathbf{t}$ by $B$ or $V$. Consider a simple dimension
$(a,b,c)=(\sqrt{1+\Delta},1,1)a_{0}$ for non-zero $|\Delta|<<1$
and $a_{0}$ is finite.  For constant time $t_{0}=0$, the
corresponding 4-momentum are
\begin{equation}
P_{\mu}=\int_{t_{0}}\mathbf{t}^{0}{}_{\mu{}ij}x^{i}x^{j}dV
=\frac{\sqrt{1+\Delta}}{12}a^{5}_{0}(\mathbf{t}^{0}{}_{\mu{}ij}\delta^{ij}+\Delta\mathbf{t}^{0}{}_{\mu{}11}).
\label{21aJan2013}
\end{equation}
Here we list out the energy component for $B$ and $V$
\begin{eqnarray}
B_{0011}&=&E_{ab}E^{ab}+H_{ab}H^{ab}-2E_{1a}E^{1a}-2H_{1a}H^{1a},\\
V_{0011}&=&3E_{ab}E^{ab}-H_{ab}H^{ab}-8E_{1a}E^{1a}+4H_{1a}H^{1a},
\end{eqnarray}
and the associated momenta are
\begin{eqnarray}
&&B_{0c11}=2\epsilon_{cab}(E^{ad}H^{b}{}_{d}-2E^{a}{}_{1}H^{b}{}_{1}),\\
&&V_{0c11}=2\epsilon_{1ab}(E^{ad}H^{b}{}_{d}-2E^{a}{}_{1}H^{b}{}_{1},
2E^{a}{}_{1}H^{b}{}_{2}-4E^{a}{}_{2}H^{b}{}_{1},
2E^{a}{}_{1}H^{b}{}_{3}-4E^{a}{}_{3}H^{b}{}_{1}).\quad\quad
\end{eqnarray}
Looking at (\ref{21aJan2013}), $\Delta\mathbf{t}^{0}{}_{\mu{}11}$
varies the energy-momentum of
$\mathbf{t}^{0}{}_{\mu{}ij}\delta^{ij}$ simultaneously. Using the
5 Petrov types Riemann curvature to compare the energy-momentum in
(\ref{21aJan2013}), we find that if $\mathbf{t}$ is replaced by
$B$ the Lorentz covariant and future directed non-spacelike
properties require $\Delta\in(-1,1]$. Similarly, if we replace
$\mathbf{t}$ by $V$, it is also true for provided
$\Delta\in[-\frac{1}{3},\frac{1}{5}]$. However, as far as the
quasilocal small 2-surface is concerned, practically, we only can
allow the non-zero $\Delta$ to be sufficiently small. Therefore,
the result in (\ref{21aJan2013}), a linear combination for
$\mathbf{t}^{0}{}_{\mu{}ij}\delta^{ij}$ with an extra
$\mathbf{t}^{0}{}_{\mu{}11}$, is a physical sensible quantities
for describing the quasilocal energy-momentum.

\section{Conclusion}
To describe the positive quasilocal energy-momentum expression,
the Bel-Robinson tensor $B$ and tensor $V$ are suitable because
both of them give the Bel-Robinson type energy-momentum in a small
cube region. In the past, people may have assumed that only this
Bel-Robinson type energy-momentum can manage this specific task:
Lorentz covariant, future pointing and non-spacelike. That
particular restriction even cannot allow any small amount of
energy to be subtracted from this Bel-Robinson type
energy-momentum. After some careful comparison and using the 5
Petrov type Riemann curvature for the verification, we have
discovered that the Bel-Robinson type energy-momentum implies
Lorentz covariant and future directed non-spacelike properties;
but the converse is not true. We find that there exists a certain
relaxation freedom such that one can (i): add an extra tensor $S$
with $B$ or $V$ in a quasilocal small cube limit, or (ii):
Directly evaluate $B$ or $V$ in a quasilocal small box region.

\section*{Acknowledgment}
The author would like to thank Dr. Peter Dobson, Professor
Emeritus, HKUST, for reading the manuscript and providing some
helpful comments. This work was supported by NSC 99-2811-M-008-021
and NSC 100-2811-M-008-063.

\end{document}